\documentclass[graybox]{svmult}

\renewcommand{\E}{{\cal{E}}}
\renewcommand{\k}{\kappa}

\makeindex             

\begin{document}

\title*{On the Properties of Exact Solutions Endowed with Negative Mass}
\author{Vladimir S. Manko}
\institute{V. S. Manko \at Departamento de F\'isica, Centro de Investigaci\'on y de
Estudios Avanzados del IPN, A.P. 14-740, 07000 M\'exico D.F., \email{vsmanko@fis.cinvestav.mx}
}
%
%
\maketitle

\abstract{It is shown that various pathological properties of spacetimes can be explained by the presence of negative mass, including the cases when the total mass of the solution is a positive quantity. As an illustration, we consider several well-known stationary axisymmetric vacuum and electrovac solutions of the Einstein-Maxwell equations. Our investigation naturally leads to a critique of the known maximal extensions of the Kerr and Kerr-Newman spacetimes which turn out to be neither analytic nor physically meaningful.}

\section{Introduction}
\label{sec:1}

The study of equilibrium configurations in the original and extended double-Kerr solutions \cite{KNe,MRu} revealed a close connection between negative mass and pathologies of a spacetime: a source carrying a negative mass necessarily develops a massless ring singularity off the symmetry axis \cite{Hoe,MRS}, even in the situations when the total mass of the equilibrium configuration is positive. Negative mass could be also responsible for the formation of the regions with closed timelike curves (CTC), as in the case of the NUT solution where negative mass has been shown to be distributed along a part of the semi-infinite massive source \cite{MRu2}. Since many other stationary axisymmetric spacetimes with positive total mass are known to have massless ring singularities outside the symmetry axis and regions with CTCs, it is of interest to find out whether these pathologies are also due to the presence of some negative mass distributions. In the present communication several recent results demonstrating the relation that exists between negative mass and space-time pathologies will be discussed. These results suggest in particular that the known maximal extensions of the Kerr and Kerr-Newman spacetimes need a critical reconsideration.

\section{Negative mass in $\delta=2$ Tomimatsu-Sato, Kerr and Kerr-Newman solutions}
\label{sec:2}
The well-known $\delta=2$ Tomimatsu-Sato (TS2) solution for a spinning mass was discovered in 1972 and is defined by an Ernst complex potential \cite{Ern} of the form \cite{TSa}
\begin{eqnarray}
\E&=&(A-B)/(A+B), \nonumber\\
A&=&p^2(x^4-1)+q^2(y^4-1)-2ipqxy(x^2-y^2), \nonumber\\
B&=&2px(x^2-1)+2iqy(y^2-1), \end{eqnarray}
where the real parameters $p$ and $q$ are subject to the constraint $p^2+q^2=1$, while the prolate spheroidal coordinates $(x,y)$ are related to the cylindrical Weyl-Papapetrou coordinates $(\rho,z)$ by the formulae
\begin{equation}
x=\frac{1}{2\k}(r_++r_-), \quad y=\frac{1}{2\k}(r_+-r_-), \quad r_\pm=\sqrt{\rho^2+(z\pm\k)^2},
\end{equation}
$\kappa$ being a positive constant.

The total mass $M_T$ of the TS2 solution is given by the expression $M_T=2\kappa/p$, so it is a positive quantity for $p>0$. However, this solution has both a massless ring singularity outside the symmetry axis \cite{TSa} and a region with CTCs \cite{GRu}. In a recent paper \cite{Man} the exact analytic formulae defining the negative mass distribution in the TS2 spacetime have been obtained with the aid of Komar integrals \cite{Kom}, and evidence relating negative mass to the pathologies of this spacetime has been provided. Moreover, the analysis of the TS2 solution with total negative mass ($p<0$) carried out in \cite{Man} shows that the case $M_T<0$ is characterized by appearance of new pathological features: the ring singularity changes its location, passing from the inner to the outer stationary limit surface (SLS), while the region with CTCs, now located outside the outer SLS, becomes visible to a distant observer. It is of interest to compare the latter case with the Kerr \cite{Ker} and Kerr-Newman \cite{New} spacetimes endowed with negative mass.

\subsection{The Kerr solution with $M<0$}
\label{subsec:2}
Due to the simple form of the Kerr solution, the corresponding negative-mass case permits a purely analytic investigation to be carried out in \cite{Man}. Quite interestingly, the ring singularity of the Kerr metric with $M<0$ lies in the equatorial plane ($z=0$) off the symmetry axis (at $\rho=|a|$, $a$ being the angular momentum per unit mass). It is massless and thus reminiscent of the case of the TS2 spacetime. The ring singularity is a locus of points where the SLS touches the region with CTCs, the latter region having a toroidal topology.

\subsection{The Kerr-Newman solution with $M<0$}
According to a recent study \cite{Man2}, the presence of an electromagnetic field in the Kerr-Newman (KN) solution slightly affects the location of the singularity developed by the negative mass compared with the vacuum Kerr case. In particular, the ergoregion and the region with CTCs do not intersect, neither do they touch each other, and the massless ring singularity is located entirely inside the latter region of the causality violation.

\section{Some remarks on the maximal extensions of the Kerr and KN spacetimes}

A thorough analysis of the Kerr and KN solutions with negative mass by Garc\'ia-Compe\'an and Manko \cite{GMa} leads to the conclusion that the known maximal analytic extensions of the stationary black-hole spacetimes proposed by Boyer and Lindquist \cite{BLi} and by Carter \cite{Car} are erroneous. As a matter of fact, the known extensions are physically inconsistent because each of them represents an artificial unification of two different spacetimes corresponding to masses of opposite signs (the region $r<0,M>0$ is similar to the region $r>0,M<0$ via the invariance of the Kerr and KN metrics under the change $r\to-r$, $M\to-M$). As a consequence, these extensions are not analytic on the disk joining the two asymptotically flat regions. A correct extension of $r$ into negative values must be accompanied by the simultaneous extension of $M$ into negative values too. Then the region $r<0,\,M<0$ will be identical (up to the change $\theta\to -\theta$) with the region $r>0,\,M>0$, and thus the gluing of two identical spacetimes on the disk encircled by the ring singularity will be analytic.

\end{document}